\documentclass{elsart}
\usepackage{natbib}
\usepackage{epsfig}

\begin{document}

\begin{frontmatter}
\title{Dynamics of Money and Income Distributions}
\author{Przemys\mbox{\l}aw Repetowicz}
\ead{repetowp@tcd.ie}
\ead[url]{www.maths.tcd.ie/$\sim$przemek}
\author{Stefan Hutzler}
\ead{shutzler@tcd.ie}
\author{Peter Richmond}
\ead{richmond@tcd.ie}
\ead[url]{www.tcd.ie/Physics/People/Peter.Richmond}
\address{Department of Physics, Trinity College Dublin 2, Ireland}
\begin{abstract}
\noindent We study the model of interacting agents proposed by \citet{Chatterjee} that allows
agents to both save and exchange wealth. Closed equations for the wealth
distribution are developed using a mean field approximation.

\noindent We show that when all agents have the same fixed savings propensity, subject
to certain well defined approximations defined in the text, these equations
yield the conjecture proposed by \citet{Chatterjee} for the form of the stationary
agent wealth distribution.

\noindent If the savings propensity for the equations is chosen according to some
random distribution we show further that the wealth distribution for large
values of wealth displays a Pareto like power law tail, ie $P(w)\sim w^{1+a}$.
However the value of $a$ for the model is exactly 1.
Exact numerical simulations for the model illustrate how, as the savings
distribution function narrows to zero, the wealth distribution changes from
a Pareto form to to an exponential function. Intermediate regions of wealth
may be approximately described by a power law with $a>1$. However the value
never reaches values of $\sim 1.6-1.7$ that characterise empirical wealth data.
This conclusion is not changed if three body agent exchange processes are
allowed.
We conclude that other mechanisms are required if the model is to agree with
empirical wealth data.
\end{abstract}

\end{frontmatter}

\section{Introduction}

The distribution of wealth or income in society has been of great interest
for many years. Italian economist \citet{Pareto} was the first to suggest it
followed a ``natural law'' where the higher end of the wealth distribution
is described by power law, $P\left( w\right) \sim w^{-1-\alpha }$. Repeated
empirical studies 
\citet{LevySolomon,Dragulescu,ReedHughes,AoyamaSouma} 
show that the power law tail exhibits a remarkable spatial
and temporal stability and while the value of the exponent, $\alpha $, may
vary slightly, it changes little from the value $\sim 1.5$. 

Even though the collected data stem from different sources
and can be incomplete because of difficulties in accessibility
(poor conclusions from income data in Sweden in \citet{LevySolomon} 
due to a too small number of wealth ranges in the data;
total net capital of individual at death in the United States (US) reported to the 
Bureau of Census and the Inland Revenue for tax heritage purposes in \citet{Dragulescu};
distributions of sizes of incomes, cities, internet files, biological taxa,
gene family and protein family frequencies in \citet{ReedHughes};
and income distributions in the Japan in \citet{AoyamaSouma})
the common conclusion which can be drawn is that
the high end that exhibits the power law is characterised by several 
multiples or even tens of multiples of the average income/wealth
(only $5\%$ of population income-data in the US conforms to a 
power-law and
the power law for the yearly income data in the United Kingdom 
sets in only for $>50 k\pounds $ \citet{Dragulescu},
income distributions in the Japan in 2000 exhibit power laws only for 
$>5\cdot 10^4$ thousands of Yen).

For around $100$ years the tantalising Pareto law remained without explanation. The
renewed interest by physicists and mathematicians in econo- and sociophysics
has however led to publication of a number of new papers on the topic in
recent years (see \citet{Slanina} for an extensive literature review).

The fact that multiplicative power law processes can lead to power law
distributions has been known for many years from studies as diverse as the
frequency of words in
text \citet{Yule}, economic growth \citet{Gibrat}, city populations %
\citet{Zipf}, wealth distribution \citet{Ijiri} and stochastic renewal
processes \citet{Kesten}.

In the analysis of these distributions \citet{Solomon} has recently 
proposed the use of Generalised Lotka
Volterra (GLV) equation that combines a multiplicative random process with
an autocatalytic process. The latter redistributes a fraction of the total money to
ensure the money possessed by an agent is never zero. This simulates in a
simplistic way the effect of a tax. The model equations lead to a wealth
distribution $P(w)$ of the form: 
\begin{equation}
P\left( w \right) \sim \frac{e^{(1-\alpha)/w}}{w^{1+\alpha}}
\end{equation}
where and $\alpha - 1$ is a positive number that is a ratio
of parameters of the model that are related to social security and some random 
investements respectively.
For large values of income $w$ this indeed exhibits a Pareto behaviour.

However
two issues arise. The first is that empirical studies of income
distributions show that this function does not describe well the very low
end of the income distribution which is essentially exponential%
\citet{Dragulescu}. The second relates to use of the multiplicative
stochastic term. It is certainly necessary to secure the right form for the
distribution function but how does it arise in the first place?

More recently \citet{Chatterjee} have developed a model of pairwise interacting
agents $i$ and $j$ that exchange money by analogy with an ensemble of gas molecules that
exchange momentum. In \citet{Chatterjee}'s model, however, the agents are allowed to save
a fraction $\lambda_i$ of their money prior to an
interaction. The total money held between 
two agents is conserved during the interaction process.
The governing equations for the evolution of wealth $w_i$ and $w_j$
of agents $i$ and $j$ respectively are given by:
\begin{eqnarray}
w_{i}\left( t+1\right) &=&\lambda _{i}w_{i}+\epsilon \left[ (1-\lambda
_{i})w_{i}+(1-\lambda _{j})w_{j}\right]  \nonumber \\
w_{j}\left( t+1\right) &=&\lambda _{j}w_{j}+\left[ 1-\epsilon \right] \left[
(1-\lambda _{i})w_{i}+(1-\lambda _{j})w_{j}\right]  \label{eq:collision}
\end{eqnarray}

Here each agent, $i$, has a savings propensity, $\lambda_i$. The remaining
money is divided during the exchange process in a random manner determined
by a uniformly distributed random number $\epsilon$ between zero and one. 
From their numerical calculations \citet{Chatterjee} found the following
results for the stationary wealth distribution $P(w)$. Here
\begin{enumerate}
\item  With no saving ($\lambda _{i}$ = 0 for all $i$) agents behave
randomly and the distribution follows the Gibbs rule $P(w)\sim exp(-w/\left<w\right>)$ 
where $\left<w\right>$ is the average wealth of agent.
$P(w)$ has a maximum when $w=0$.

\item  If the saving propensity is non-zero and takes the same constant value for
all agents ($\lambda _{i}=\lambda $ for all $i$) the resulting distribution
can be fitted well by the heuristic function: 
\begin{equation}
P\left( w\right) =\frac{n^{n}}{\Gamma (n)}w^{n-1}\exp \left( -nw\right) 
\label{eq:Distrib}
\end{equation}
where $\Gamma(n)$ is the gamma function and
the parameter $n$ is related to the saving propensity, $\lambda $ as
follows: 
\begin{equation}
n\left( \lambda \right) =1+\frac{3\lambda }{1-\lambda }  \label{eq:Relat}
\end{equation}
The power, $w^{n-1}$, qualitatively changes the distribution so that it has
a maximum for $w>0$.
The author does not give any theoretical arguments for the use of this distribution.

\item  If the saving propensity for the agents is chosen according to some
random distribution, like uniform or power-law distributions with $0 \le \lambda \le 1$,
the numerical output for large values of money gives $%
P\left( w\right) \sim w^{-1-\alpha }$. This is the celebrated Pareto law.
Numerical calculations yield a value for $\alpha =1.03\pm 0.03.$ The authors
show empirical data for wealth distributions for both Japan and the USA.
These data clearly exhibit power laws with values of $\alpha $ greater than
this value. However the authors leave the reader wondering whether the model
could fit this data better. A further calculation allowing only a fraction, $%
p$, of agents to save is made but the the value for the Pareto law remains
unchanged. The authors do not investigate the possible changes in $\alpha $
as a result of using savings distributions that differ from uniform
distributions.
\end{enumerate}

This work is interesting in that it brings together within one framework the
distributions of both Gibbs and Pareto. However it leaves open tantalizing
questions.

\begin{enumerate}
\item  Is it possible to predict analytically the expressions \ref
{eq:Distrib} and \ref{eq:Relat}?

\item  How does the value of $\alpha $ within the model depend on the nature
of the savings distribution?

\item  Could it be that the value of $\alpha $ is actually unity?

\item  Is there a way of reconciling the approach based on the GLV equations
and the exchange theory of Chakrabarti?
\end{enumerate}

Note that caution has to be taken by fitting the models described above to
income data obtained from Inland Revenues  in different countries.
As pointed out in \citet{DragulescuStatMech} the wealth 
has to be understood as a commodity that is subject to an incessant process of exchange
rather then as valueables like precious metals, ``hard currency'',
bonds or works of art that have been deposited in a bank account in order to serve as
a lifetime security. 
In this sense the distributions that come out of the models
should be fitted to a momentary
distribution of money in the society; a distribution they may or may not
be in equillibrium. 
Since people rarely disclose their momentary wealth 
the statistical 
data one avails of regards more the total wealth of individuals, ie the wealth that has
been accumulated throughout their whole lifetimes and is reported to the 
Revenue office only after death (to fulfill the heritage tax requirements). 
Since, however, individuals with small and medium wealths
are rather unlikely to invest parts of their income in any sort of lifetime
securities, because their earnings are small and are spent in their total 
to cover the cost of living, the low end of the momentary money distribution
in equillibrium should coincide with the low end of the wealth distribution
obtained from the Revenue data. Differences will only be observed in the high end.

In the next section we develop the theory for the model by \citet{Chatterjee} 
and show that it is
indeed possible to demonstrate that the conjecture summarised above is, to
within a certain well defined approximation, correct.
We then study asymptotically the behaviour of the wealth distribution for
the case where the savings propensity varies for the agents and demonstrate
that if the wealth distribution $P(w)\sim w^{1+\alpha}$ then $\alpha$ is exactly unity for
this model. We demonstrate in section \ref{sec:power-law} that this conclusion remains
unchanged even when three agent exchange processes are allowed.
The conclusions of the mean field analytic analysis are supported by exact
numerical simulations shown 
in Figs. \ref{fig:Exponent3Agents} and \ref{fig:Simulation1}.

\section{Theoretical analysis}

Complete information about the processes at time t is given by the N agent
distribution function $f_{N}(v_{1},\ldots ,v_{N}).$ In what follows we shall
assume the mean-field approximation. This implies that the $N$-agent
distribution function: 
\begin{equation}
f_{N}(v_{1},\ldots ,v_{N})=P\left( \bigcap_{i}v_{i}\leq V_{i}\leq
v_{i}+dv\right) /(dv)^{N}=\prod_{i}f_{1}(v_{i})
\end{equation}
We can now invoke the Boltzmann equation \citet{Ernst} for the one-agent
wealth $v$ distribution $f_{1}(v;t)$ at time $t$. Thus: 
\begin{equation}
\partial _{t}f_{1}(v;t)=\int dwdv^{\prime }dw^{\prime }\left( W\left(
v\;w|v^{\prime }\;w^{\prime }\right) f_{1}(v^{\prime })f_{1}(w^{\prime
})-W\left( v^{\prime }\;w^{\prime }|v\;w\right) f_{1}(v)f_{1}(w)\right)
\end{equation}
where the transition probabilities $W\left( {}\right) $ are given via the
rules for the collision-dynamics (\ref{eq:collision}): 
\begin{eqnarray}
\lefteqn{
W\left( v^{\prime }\;w^{\prime }|v\;w\right) =}  \nonumber \\
&&\delta \left( v^{\prime }-\left( \lambda v+\epsilon (1-\lambda )(v+w)\right)
\right) \cdot \delta \left( w^{\prime }-\left( \lambda w+(1-\epsilon
)(1-\lambda )(v+w)\right) \right) 
\end{eqnarray}
Introducing the Laplace transform $\tilde{f}_{1}(x;t):=\int_{0}^{\infty
}f_{1}(v;t)\exp (-vx)dv$ we obtain an integro-differential equation for the
temporal evolution: 
\begin{equation}
\partial _{t}\tilde{f}_{1}(x;t)+\tilde{f}_{1}(x;t)=\left\langle \tilde{f}%
_{2}\left( \lambda x+\epsilon \left( 1-\lambda \right) x,\epsilon \left(
1-\lambda \right) x;t\right) \right\rangle  \label{eq:masterI}
\end{equation}
where the random spatial variability of the saving propensities $\lambda $
and exchange fractions $\epsilon $ is accounted for by the averaging process 
$\left\langle {}\right\rangle $ over their random distributions.

We note at this point that this model assumes elastic scattering, i.e.
conservation of wealth during the exchange process (\ref{eq:collision}) and
the existence of a stationary solution. This is in contrast to many previous
models formulated in different contexts where 'wealth' may be either lost %
\citet{Krapivsky,Ben-Avraham,Bobylev,Baldassarri} or gained \citet{Slanina}
in the exchange process and the distribution function has a power-law-tail
only in an asymptotic sense.

We now write the stationary solutions $\tilde{f}_{1}(x)=lim_{t%
\longrightarrow \infty }\tilde{f}_{1}(x;t)$ both in terms of solutions of
non-linear integral equations (Master Equations (MEs)) and in terms of
expansions $\tilde{f}_{1}(x)=\sum_{n=0}^{\infty }(-1)^{n}m_{n}x^{n}$ over
moments (Ms) $\left\langle v^{n}\right\rangle =m_{n}\cdot n!$ which
satisfy recursion relations. It is convenient to distinguish two cases:


(I) Saving propensity: $\lambda \neq 0$ but equal for all agents: 
\begin{eqnarray}
\lefteqn{\mbox{ME:}\quad x\tilde{f}_{1}(x)=\frac{1}{1-\lambda }%
\int_{0}^{(1-\lambda )x}\tilde{f}_{1}(\lambda x+\phi )\tilde{f}_{1}(\phi
)d\phi }  \label{eq:MEI} \\
\lefteqn{\mbox{Ms:}\quad m_{p}=\sum_{q=0}^{p}m_{q}m_{p-q}\tilde{C}%
_{q}^{(p)}(\lambda )\quad \mbox{with}\quad \tilde{C}_{q}^{(p)}(\lambda )=%
\frac{\int_{0}^{(1-\lambda )}\left( \lambda +\eta \right) ^{q}\eta
^{p-q}d\eta }{1-\lambda }}  \label{eq:Moments} \\
\lefteqn{\mbox{and}\quad \tilde{C}_{q+1}^{(p)}=\frac{(1-\lambda
)^{p-q-1}-(q+1)\tilde{C}_{q}^{(p)}}{p-q}\quad \mbox{with}\quad \tilde{C}%
_{0}^{(p)}=\frac{(1-\lambda )^{p}}{p+1}}
\end{eqnarray}

(II) Random saving propensity: $\lambda \sim \rho _{\Lambda }(\lambda )$. 
\[
\lefteqn{\mbox{ME:}\quad x\tilde{f}_{1}(x)=\int_{0}^{1}d\lambda \frac{\rho
_{\Lambda }(\lambda )}{1-\lambda }\int_{0}^{(1-\lambda )x}\tilde{f}%
_{1}(\lambda x+\phi )\tilde{f}_{1}(\phi )d\phi } 
\]
Here we waive the writing of equations for the moments since due to to the
wealth distribution having a power-law tail they may not of course exist.
Now an assumption about an asymptotic expansion in the ``wealth-domain'': $%
f_{1}(v)=\sum_{n=0}^{\infty }a_{n}/v^{n+\alpha +1}$ leads to a decomposition
of the function in the Laplace domain into two parts 
\begin{equation}
\tilde{f}_{1}(x)=\tilde{f}_{1}^{\mbox{reg}}(x)+\tilde{f}_{1}^{\mbox{sing}}(x)
\label{eq:decomp}
\end{equation}
with the first part being an analytic function $\tilde{f}_{1}^{\mbox{reg}%
}(x)=1-x+O\left( x^{2}\right) $ and the second part $\tilde{f}_{1}^{%
\mbox{sing}}(x)=\sum_{n=0}^{\infty }b_{n}x^{n+\alpha }$ having a leading
term $x^{\alpha }$ of order $\alpha $.

\subsection{\noindent Conjecture by Patriarca, Chakraborti and Kaski (PCK):}

\begin{figure}[tbp]
\centerline{\psfig{figure=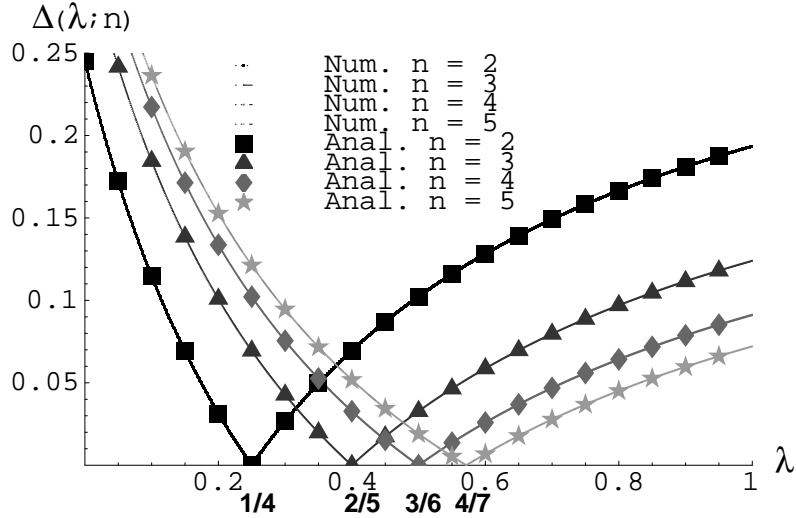,width=0.8\textwidth,angle=0}}
\caption{Deviations $\Delta (\protect\lambda ;n)=\sum_{p=0}^{10}\left|
(m_{p}(\protect\lambda )-m_{p}^{{conj}}(n))/p!\right| $ of the exact moments $%
m_{p}$ of the wealth distribution from the moments $m_{p}^{{\mbox{conj}}}$ derived from
the conjecture plotted as a function of $\protect\lambda $ for $n=2,3\ldots
,9$. Solid lines (dot symbols) correspond to analytical (numerical) 
solutions of the moment equations 
(\ref{eq:Moments}).
We see that the minima $\protect\lambda =(n-1)/(n+2)$ $=$ $\left\{
1/4,2/5,3/6,4/7,5/8,6/9,7/10,8/11\right\} $ of the deviations do correspond
to the PCK conjecture (\ref{eq:Relat}).
\label{fig:MomentDeviations}}
\end{figure}

Solving the moments' equations (\ref{eq:Moments}) with initial conditions $%
m_{0}=1$ and $m_{1}=1$ recursively, ie. expressing, via the $p$th equation, $%
m_{p}$ as a function of $\lambda $ and all previous values of $m$, (ie $%
m_{0},m_{1},\ldots ,m_{p-1}$), one obtains: 
\begin{eqnarray}
m_{2} &=&\frac{\lambda +2}{2(1+2\lambda )}\quad m_{3}=\frac{\lambda +2}{%
2(1+2\lambda )^{2}} \\
m_{4} &=&\frac{72+12\lambda -2\lambda ^{2}+9\lambda ^{3}-\lambda ^{5}}{%
24(1+2\lambda )^{2}(3+6\lambda -\lambda ^{2}+2\lambda ^{3})}
\label{eq:ChakrabartiMoments}
\end{eqnarray}
The first three moments $m_{1},m_{2}$ and $m_{3}$ coincide with the moments
of PCKs function (\ref{eq:Distrib}) if the relation between the parameters $%
n $ and $\lambda $ is given by (\ref{eq:Relat}). Indeed the coefficients of
a series expansion of the Laplace transform 
\begin{equation}
\tilde{P}(x)=\int_{0}^{\infty }P(\xi)\exp (-\xi x)d\xi=\left( \frac{n}{x+n}\right)
^{n}
\end{equation}
of the function (\ref{eq:Distrib}) agree with moments (\ref
{eq:ChakrabartiMoments}) up to the third order subject to equation (\ref
{eq:Relat}) being satisfied. This is shown in a nice way in Fig. \ref{fig:MomentDeviations}.
The deviation $\Delta \tilde{f}_{1}(x)$ between
the exact solution of the ME (\ref{eq:MEI}) and the ansatz (\ref{eq:Distrib}%
) has a leading fourth order: 
\begin{equation}
\Delta \tilde{f}_{1}(x)=\frac{(n-1)(n+1)(n+8)}{8n^{3}(10n^{3}+30n^{2}+45n-4)}%
x^{4}+O\left( x^{5}\right)
\end{equation}
It is hard to say if a more general class of functions than (\ref{eq:Distrib}%
) would satisfy the ME to higher expansion orders.

\subsection{\noindent The power-law tail: \label{sec:power-law}}

Calculations by \citet{Chatterjee} and ourselves suggest that the
value of the exponent $\alpha $ is equal to one. Let us look at this aspect
in more detail. Inserting the expansion (\ref{eq:decomp}) into the ME (\ref
{eq:MEI}) and comparing coefficients of order $x^{\alpha +1}$ on both sides
of the equation leads to a transcendental equation for the exponent, $\alpha 
$: 
\begin{equation}
\left\langle (1-\lambda )^{\alpha }\right\rangle +\left\langle \frac{%
1-\lambda ^{\alpha +1}}{1-\lambda }\right\rangle =\alpha +1
\label{eq:Exponent}
\end{equation}
Clearly if we choose $\alpha =1$, (\ref{eq:Exponent}) we obtain an indentity
for any distribution of $\lambda $. This would seem to be true even for a
distribution that assumes only a fraction $p$ of the agents save and the
remainder do not save, i.e. $\rho _{\Lambda }^{(1)}(\lambda )=p\rho
_{\Lambda }(\lambda )+(1-p)\delta (\lambda )$. 

However, whether other
solutions for $\alpha $ exist is an open question and depends on the
distribution of the saving propensity $\rho _{\Lambda }(\lambda )$. 
We try to clarify this question below.

\begin{figure}[tbp]
\centerline{\psfig{figure=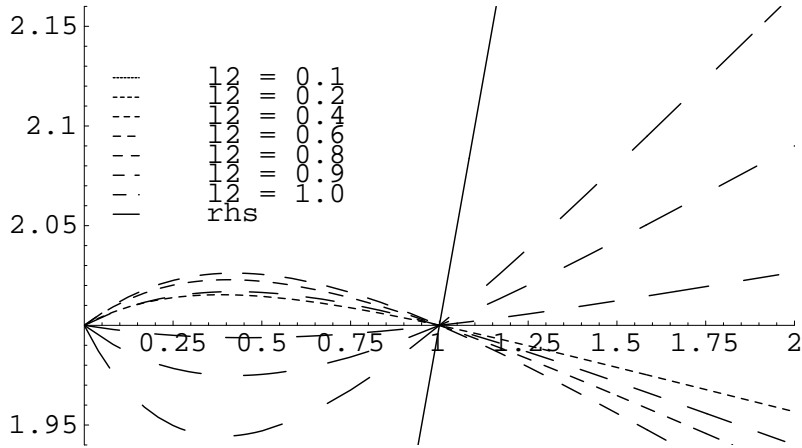,width=0.8\textwidth,angle=0}}
\caption{The left-hand-side of equation (\ref{eq:Exponent}) plotted as a
function of $\protect\alpha $ for a uniformly random saving propensity $%
\protect\lambda $ with $0<\protect\lambda <l_{2}$ for different values of $%
l_{2}$ (dashed lines) and the right-hand-side (rhs) of the equation (\ref
{eq:Exponent}) (solid line). As we can see there is no other solution
of the transcendental equation (\ref{eq:Exponent}) 
except $\alpha=1$ in the range $\alpha \in [0,2]$.}
\label{fig:Exponent}
\end{figure}

\begin{figure}[tbp]
\centerline{\psfig{figure=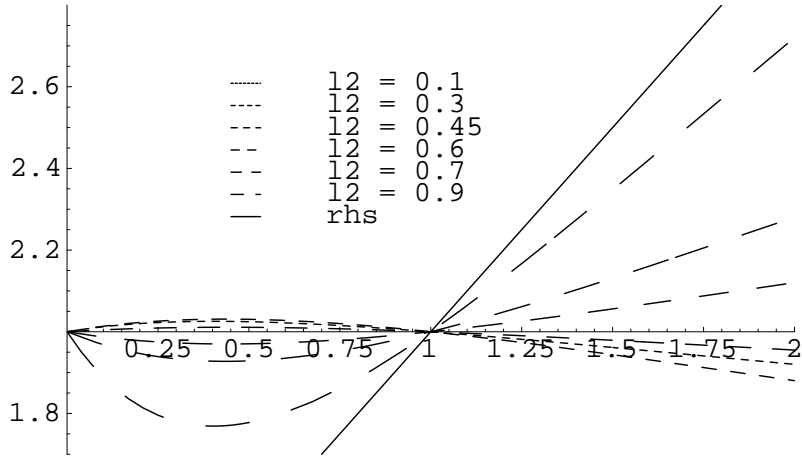,width=0.8\textwidth,angle=0}}
\caption{The same as in Fig.\ref{fig:Exponent} but for the propensity $%
\protect\lambda $ conforming to a truncated $\left| \protect\lambda \right|
\leq 1$ normal distribution with variable mean $l_{2}$ and standard
deviation $0.01$. As before dashed lines denote the left-hand side and the
solid line denotes the right-hand side of equation (\ref{eq:Exponent}).
Again the only solution of the transcendental equation is $\alpha=1$
in the range $\alpha \in [0,2]$. }
\label{fig:ExponentGauss}
\end{figure}

\begin{figure}[tbp]
\centerline{\psfig{figure=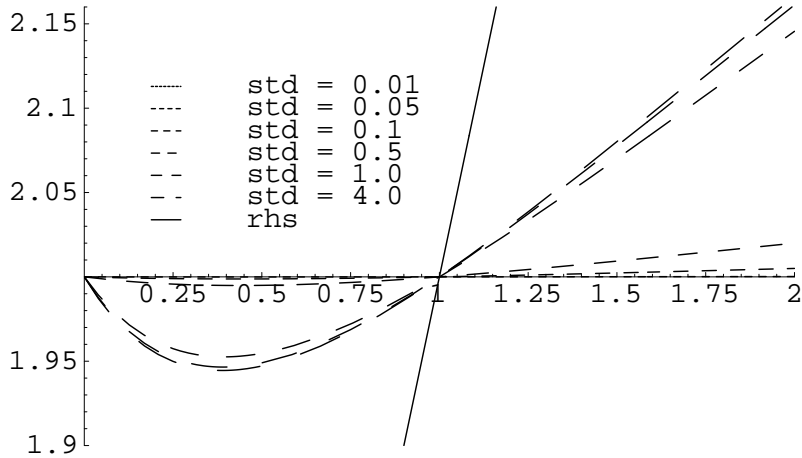,width=0.8\textwidth,angle=0}}
\caption{The same as in Figs.\ref{fig:Exponent} and \ref{fig:ExponentGauss}
except that now the mean of the normal distribution of propensities $\protect%
\lambda $ is fixed and equal to $0.5$ and the standard deviation varies.
Here again no new solutions except $\alpha=1$ are obtained.}
\label{fig:ExponentGaussII}
\end{figure}

For uniformly distributed propensities $\rho _{\Lambda }(\lambda )=1/l_{2}$ for $%
0\leq \lambda \leq l_{2}\leq 1$ the only solution is $\alpha =1$ (see Fig. 
\ref{fig:Exponent}). Likewise if $\rho _{\Lambda }(\lambda )$ is a normal
distribution with a variable mean $l_{2}$ where $\left| l_{2}\right| \leq 1$
and the standard deviation is small (See Fig.\ref{fig:ExponentGauss}) or
with a fixed mean and variable standard deviation (Fig.\ref
{fig:ExponentGaussII}) the exponent similarly turns out to be unity.

Now we make a stronger statement and say that there is no continuous
and differentiable distribution of saving propensities $\lambda \in [0,1]$ 
that would yield $\alpha \ne 1$.   
Indeed since every distribution $\rho _{\Lambda }(\lambda )$ can be 
constructed as a weighted (possibly continuous) linear combination 
of uniform distributions
\begin{equation}
\rho_{\Lambda }(\lambda ) = \int_0^1 w(\nu) U(0, \nu) d\nu
\end{equation}   
and since for a uniform distribution $U(0, \nu)$
the left-hand side of the transcendental equation (\ref{eq:Exponent}) 
intersects the right-hand side only for $\alpha=1$ in the range 
$\alpha \in [0,2]$ then the last statement holds also for a generic distribution 
$\rho_{\Lambda }(\lambda )$. 
Here we used conditional averaging. This means that 
the average in equation (\ref{eq:Exponent}) 
is carried out as an average over $U(0, \nu)$ conditioned on $\nu$ first
and then over the distribution $w(\nu)$ of random values of $\nu$. 
 


\subsection{Beyond the mean-field approximation:}

Many-agent distribution functions $f_N(x_1,\ldots,x_N)$ may not be produced
correctly within the mean-field approach. Furthermore the wealth-exchange
model by \citet{Chatterjee} may be extended to N-point interactions: 
\begin{equation}
w_i\left(t+1\right) = \lambda_i w_i + \epsilon_i \left[\sum_{j=1}^N
(1-\lambda_j) w_j\right] \quad \sum_i \epsilon_i = 1
\label{eq:Npointcollision}
\end{equation}
Here we mean that at every time step exchange processes involving any number
of agents can happen -- each with a certain likelihood. We perform the
analysis for $N=3$ in order to find out what kind of mathematical
difficulties we will come across. Now the master equation for the 2-agent
distribution function in the Laplace domain (compare with (\ref{eq:masterI}%
)) reads:

\newpage 
\begin{eqnarray}
\lefteqn{ \partial_t \tilde{f}_2(x,y;t) + \tilde{f}_2(x,y;t) = }  \nonumber
\\
&&(1 - \sigma) \left<\tilde{f}_2 \left( \lambda x + \lambda_1 \left(\epsilon x +
(1-\epsilon) y \right), \lambda y + \lambda_1 \left(\epsilon x +
(1-\epsilon) y \right) ; t\right) \right> +  \nonumber \\
&& \frac{\sigma}{2} \left< \tilde{f}_3\left( \lambda x + \lambda_1
\left(\epsilon_1 x + \epsilon_2 y \right), \lambda y + \lambda_1
\left(\epsilon_1 x + \epsilon_2 y \right), \lambda_1 \left(\epsilon_1 x +
\epsilon_2 y \right) ; t\right) \right> +  \nonumber \\
&& \frac{\sigma}{2} \left< \tilde{f}_3\left( \lambda_1 \left(\epsilon_2 x +
\epsilon_3 y \right), \lambda x + \lambda_1 \left(\epsilon_2 x + \epsilon_3
y \right), \lambda y + \lambda_1 \left(\epsilon_2 x + \epsilon_3 y \right) ;
t\right) \right>  \label{eq:masterII}
\end{eqnarray}

where $\lambda + \lambda_1 = 1$, $\epsilon + \epsilon_1 \le 1$ and $\sigma$ 
and ($1 - \sigma$)
denote likelihoods of three-agent and two-agent exchange processes respectively. 
The first (second and third) term(s) on the
right-hand side in (\ref{eq:masterII}) account(s) for two-(three-)agent
exchange processes repectively. Setting $y=0$ we obtain equation (\ref
{eq:masterI}) except for three-agent exchange terms that were neglected in
the first place and now have been added appropriately. Setting $x=y=0$ we
obtain an identity from the normalisation condition $f_2(0,0) = f_3(0,0,0) = 1$.

\subsection{\noindent The power-law tail with three-agent exchange processes:}
\begin{figure}[tbp]
\centerline{\psfig{figure=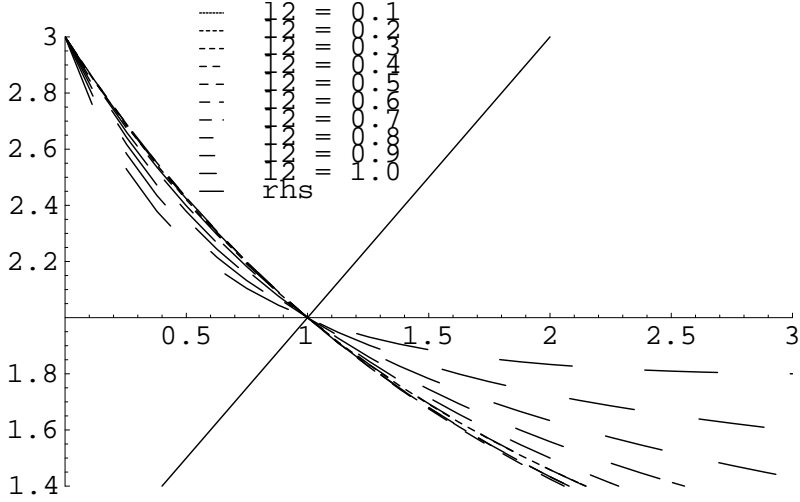,width=0.8\textwidth,angle=0}}
\caption{The second term (corresponding to three agent exchange processes) 
on the left-hand-side of equation (\ref{eq:ExponentThreeAgents}) plotted as a
function of $\protect\alpha $ for a uniformly random saving propensity $%
\protect\lambda $ with $0<\protect\lambda <l_{2}$ for different values of $%
l_{2}$ (dashed lines) and the right-hand-side (rhs) of the equation (\ref
{eq:Exponent}) (solid line). We see that three-agent exchange processes
do not lead to a change of the exponent $\alpha$ from unity to a different value.}
\label{fig:Exponent3Agents}
\end{figure}

Now the transcendental equation, derived from the master equation (\ref{eq:masterII}),
has the following form:
\begin{eqnarray}
\lefteqn{
(1 - \sigma) \left[
\left\langle (1-\lambda )^{\alpha }\right\rangle +\left\langle \frac{%
1-\lambda ^{\alpha +1}}{1-\lambda }\right\rangle 
\right]
+} \nonumber \\
&&\sigma \frac{2}{\alpha + 2}
\left[ 2 \left\langle  (1-\lambda)^{\alpha} \right\rangle  +
         \left\langle \frac{1 - (\alpha+2)\lambda^{\alpha+1} + (\alpha+1)\lambda^{\alpha+2}}{(1 - \lambda)^2} \right\rangle 
\right]
=\alpha +1
\label{eq:ExponentThreeAgents}
\end{eqnarray}
and a $\alpha=1$ is again the only solution (compare Fig.\ref{fig:Exponent3Agents})
of this equation for arbitrary 
saving propensity distributions
$\rho _{\Lambda }(\lambda )$
and for any likelihood $\sigma \in [0,1]$ of three-agent exchange processes.
This is in conformance with our numerical simulations that also show that an introduction of three-agent exchange processes do not alter the exponent.

\subsection{Moment equations in the case of two- and three-agent exchange processes:}

The expansion of the steady-state solution in terms of two-agent
correlations $\left< v^p w^q \right> =  m_{p,q} \cdot (p! q!)$ now
reads $f_2(x, y) = \sum_{p,q=0}^\infty (-1)^{p+q} m_{p,q} x^p y^q$ and the
two-agent correlations satisfy following equations: 
\begin{eqnarray}
\lefteqn{ m_{p,q} = (1 - \sigma) \sum\limits_{{\tiny \begin{array}{c} p_1 + q_1 = p\\p_2
+ q_2 =q\end{array}}} \left( \begin{array}{c} p_1 + p_2 \\ p_1 \end{array}
\right) \left( \begin{array}{c} q_1 + q_2 \\ q_1 \end{array} \right)
m_{p_1+p_2, q_1+q_2} \tilde{C}_{p_1,q_2}^{(p,q)}(\lambda) + }  \nonumber \\
&&\sigma \sum\limits_{{\tiny 
\begin{array}{c}
p_1 + q_1 +r_1=p \\ 
p_2 + q_2 + r_2 = q
\end{array}
}} \left( 
\begin{array}{c}
p_1 + p_2 \\ 
p_1
\end{array}
\right) \left( 
\begin{array}{c}
q_1 + q_2 \\ 
q_1
\end{array}
\right) \left( 
\begin{array}{c}
r_1 + r_2 \\ 
r_1
\end{array}
\right) m_{p_1+p_2,q_1+q_2,r_1+r_2} \tilde{D}_{p_1,q_2}^{(p,q)}(\lambda)
\end{eqnarray}
where 
\begin{eqnarray}
\tilde{C}_{p_1,q_1}^{(p,q)}(\lambda) &=& \frac{1}{\lambda}
\int_0^{\lambda_1} d\xi \left(\lambda + \xi\right)^{p_1} \xi^{p-p_1} \left(1
- \xi\right)^{q_1} \left(\lambda_1 - \xi\right)^{q-q_1}  \nonumber \\
\tilde{D}_{p_1,q_1}^{(p,q)}(\lambda) &=& \frac{2}{\lambda_1^2}
\int_0^{\lambda_1} d\xi \int_0^{\lambda_1-\xi} d\eta \left(\lambda +
\xi\right)^{p_1} \xi^{p-p_1} \left(\lambda + \eta\right)^{q_1} \eta^{q-q1}
\label{eq:Integrals}
\end{eqnarray}
If we neglect ternary exchange processes ($\sigma=0$) we end up with
following equations for the moments of the 1-agent distribution function: 
\begin{eqnarray}
\lefteqn{ m_2 = m_{1,1}\frac{\lambda + 2 }{2 (1 + 2 \lambda)} \quad m_3 = 
\frac{m_{1, 2}}{1 + 2 \lambda}}  \nonumber \\
&&m_4 = m_{1,3} \frac{(4 - 2 \lambda + 2 \lambda^2 + \lambda^3)} {2 (3 + 6
\lambda - \lambda^2 + 2 \lambda^3)} - m_{2,2} \frac{-6 + 3 \lambda + 2
\lambda^2 + \lambda^3} {6 (3 + 6 \lambda - \lambda^2 + 2 \lambda^3)} 
\nonumber \\
&&m_5 = \frac{((2 - 2 \lambda + 3 \lambda^2) m_{1, 4} - 2 (-1 + \lambda)
m_{2, 3})}{ (4 + 8 \lambda - 3 \lambda^2 + 6 \lambda^3)}
\label{eq:ChakrabartiMomentsII}
\end{eqnarray}
These equation reduce to equations (\ref{eq:ChakrabartiMoments}) under the
mean-field approximation $m_{p,q} = m_p m_q$.

\section{Exact computer simulations of the model}

Intuitively, it is clear that the distribution function depends critically
on the relative values of the mean value of the savings propensity and the
spread or mean square deviation of the saving distribution function. As the
spread of saving propensities tends to zero, the distribution function must
change from one having a power law tail to one with an exponential tail. How
does this change take place? Could it be that the effective power law region
shifts to take on values of $\alpha >1$? Might this explain the empirically
observed facts? We have made computer simulations of the model for different
values of these parameters. The results are described below.

\begin{figure}[tbp]
\centerline{\psfig{figure=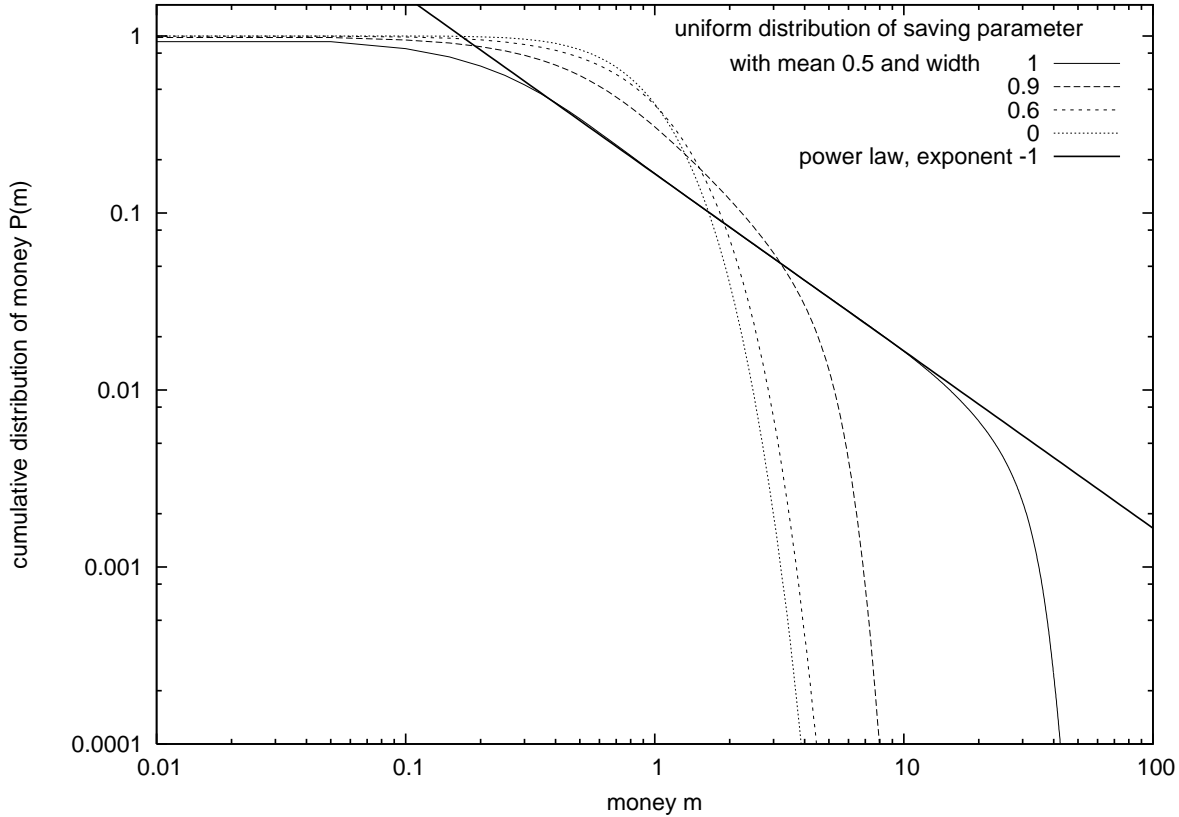,width=0.8\textwidth,angle=-90}}
\caption{
Results of computer simulations, 
obtained for
500 agents, and 39000 realisations, each taken after 100000 equilibration
steps. The data is well described by a power law with exponent -1 in the
range of 0.3 up to about 15 times the average money per agent. Changing
the width of the distribution of $\lambda$ immediately leads to a loss of
the power law. The latter may be a finite size effect since our analytical calculations
from section \ref{sec:power-law}  show that the power-law 
does occur for uniform distributions of any widths. 
\label{fig:Simulation1}}
\end{figure}

A uniform distribution of the savings parameter $\lambda$ in the model of
\citet{Chatterjee} results in a power law distribution of the cumulative
distribution shown in the Fig. \ref{fig:Simulation1}.

\begin{figure}[tbp]
\centerline{\psfig{figure=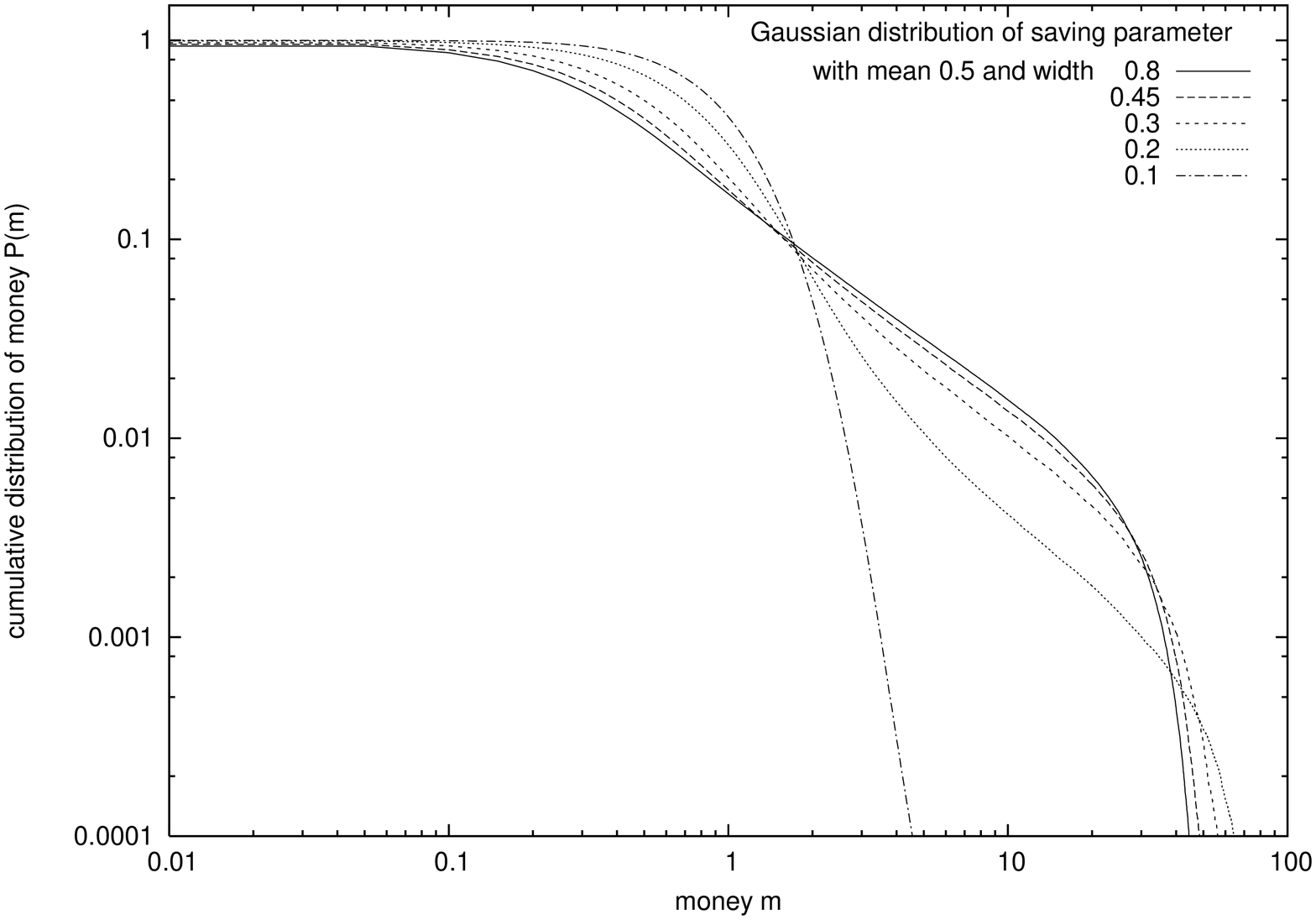,width=0.8\textwidth,angle=-90}}
\caption{
The same as in Fig. \ref{fig:Simulation1} but for the saving propensity being
Gaussian distributed.
Here the power law exponent decreases with decreasing width, 
but only over a small range from
-1 to -1.13. Narrower Gaussian distributions do not result in wealth
distributions that can be described by power law. (Note that in the limit
of small widths both Gaussian and uniform distribution give the same
wealth distribution.)
\label{fig:Simulation2}}
\end{figure}

The situation is different in the case of $\lambda$ being Gaussian
distributed. Here the cumulative wealth distribution may still be
approximately described by a power law for widths between 1 and 0.45
(see Fig. \ref{fig:Simulation2}).



One thing seems clear. The region where power law behaviour is observed is
fairly well marked even when the spread is small. And moreover the slope is
consistent with higher values of $\alpha .$ However it does not seem that $%
\alpha $ can take on values greater than around 1.2. As it stands then we
conclude that this model does not offer a complete picture of the
empirically observed wealth dynamics.

\section{Conclusions}
We have studied the model of interacting agents proposed by \citet{Chatterjee} that
allows agents to both save and exchange wealth. Closed equations for the
wealth distribution are developed using a mean field approximation.

We have shown that when all agents have the same fixed savings propensity,
subject to certain well defined approximations defined in the text, these
equations yield the conjecture proposed by \citet{Chatterjee}  for the form of the
stationary agent wealth distribution.

If the savings propensity for the equations is chosen according to some
random distribution we have further shown that the wealth distribution for
large values of wealth displays a Pareto like power law tail, ie
$P(w)\sim w^{1+a}$. However the value of a for the model is exactly 1.
Exact numerical simulations for the model illustrate how, as the savings
distribution function narrows to zero, the wealth distribution changes from
a Pareto form to to an exponential function. Intermediate regions of wealth
may be approximately described by a power law with $a>1$. However the value
never reaches values of $\sim 1.6-1.7$ that characterise empirical wealth data.
This conclusion is not changed if three body agent exchange processes are
allowed.
We conclude that other mechanisms are required if the model is to agree with
empirical wealth data.

\end{document}